\documentclass{aa}    
\def\ut#1{\mathop{\vtop{\ialign{##\crcr
     $\hfil\displaystyle{#1}\hfil$\crcr\noalign
     {\kern1pt\nointerlineskip}\hbox{$\hfil\sim\hfil$}\crcr
     \noalign{\kern1pt}}}}}

\def\undersymbol#1#2{\mathop{\vtop{\ialign{##\crcr
     $\hfil\displaystyle{#2}\hfil$\crcr\noalign
     {\kern1pt\nointerlineskip}\hbox{$\hfil#1\hfil$}\crcr
     \noalign{\kern1pt}}}}}

\def\arcmin{^{\prime}}

\begin{document}
\title{EUV excess in the inner Virgo cluster}
\author{
    F. De Paolis,
    G. Ingrosso,
  A. A. Nucita
   and
     D. Orlando
     }
\institute{Dipartimento di Fisica, Universit\`a di Lecce and INFN,
Sezione di Lecce, Via Arnesano CP 193, I-73100 Lecce, Italy }
\offprints{F. De Paolis, e-mail: depaolis@le.infn.it}
\date{Received 24 October 2001 / Accepted 21 October 2002}
\authorrunning{F. De Paolis et al.}
\titlerunning{EUV excess in the inner Virgo cluster}
\maketitle
\begin{abstract}

Observations with the Extreme UltraViolet Explorer (EUVE)
satellite have shown that the inner region of the Virgo cluster
(centered in M87 galaxy) has a strong Extreme UltraViolet (EUV)
emission (up to $13^{\prime}$) in excess to the low-energy tail
expected from the hot, diffuse IntraCluster Medium (ICM).
Detailed observations of large scale radio emission and upper
limits for hard, non-thermal X-ray emission in the $2 - 10$ keV
energy band have been also reported. Here we show that all
available observations can be accounted for by the existence of
two electron Populations (indicated as I and II) in the M87
Galaxy. The mildly relativistic Population I is responsible for
the EUV excess emission via IC scattering of CBR and starlight
photons. Population II electrons (with higher energy) are instead
responsible for the radio emission through synchrotron mechanism.
The same electrons also give rise to hard non-thermal X-ray
emission (via IC scattering of CBR photons), but the resulting
power is always below the upper bounds placed by present
observations. The non-negligible energy budget of the two electron
populations with respect to that associated with thermal
electrons indicates that the M87 galaxy is not today in a
quiescent (relaxed) phase. Nuclear activity and merging processes
could have made available this energy budget that today is
released in the form of relativistic electrons.


\end{abstract}

\section{Introduction}

Observations with the EUVE satellite (Bowyer \& Malina
\cite{euve}) have provided evidence that a number of clusters of
galaxies produce intense EUV emission, substantially in excess of
the low-energy tail expected from the X-ray emission by the hot,
diffuse ICM at temperature of a few keV. EUV excesses have been
reported for Virgo (Lieu et al. \cite{virgo}), Coma (Lieu et al.
\cite{coma}), A1795 (Mittaz, Lieu \& Lockman \cite{a1795}), A2199
galaxy clusters (Lieu, Bonamente \& Mittaz \cite{a2199}), A4038
(Bowyer, Lieu \& Mittaz \cite{blm98}) and A4059 (Durret et al.
\cite{dsldb}). However, although there is no doubt about the
detection of the EUV excess emission for Virgo and Coma clusters,
considerable controversy  still exists for the other clusters
since EUVE results are affected by the variation of the telescope
sensitivity over the field of view and depend also on subtraction
of background signals that do not come from the cluster (see
Bergh\"{o}fer \& Bowyer \cite{bb2002} and Durret et al.
\cite{dsldb}).

The initial explanation for the EUV excess emission was that it
is produced by a warm ($T \sim 10^6$ K) gas component of the ICM
(Lieu et al. \cite{coma}). For Virgo and Coma clusters,
Bonamente, Lieu \& Mittaz (\cite{blm2001}) have proposed that HI
cold clouds might have, at least in part, been released in the
intergalactic space by ram pressure stripping. In this framework,
warm gas can be generated at the interface between the cold phase
(HI clouds) and the hot ICM, e.g. by the {\it mixing layer}
mechanism (Fabian \cite{f97}). The resulting scenario is then a
three phase gas model, where the hot and warm gas components are
responsible for the bulk of X-ray emission and the soft excess
emission, respectively, while the cold component accounts for
X-ray absorption (Buote \cite{b2001}).

Models invoking the presence in galaxy clusters of the warm gas
component have problems since they refer to unrelaxed or
turbulent gas conditions: the longevity of such multiphase gas
would then require the presence of a heat source to carefully
balance the radiative cooling, but no obvious heating mechanisms
have been found (Fabian \cite{f96}). Indeed, it is difficult for
a warm intracluster gas to remain for a long time at $\sim 10^6$K
since at this temperature the cooling function for a gas with
solar chemical composition has a value $\Lambda\simeq 2\times
10^{-22}$ erg cm$^3$ s$^{-1}$ (Saito \& Shigeyama \cite{ss})
implying for the M87 galaxy warm gas component with metallicity
$Z\simeq 0.2-0.3 ~Z_{\odot}$ (Gastaldello \& Molendi
\cite{molendi}) and mean number density $n\simeq 10^{-3}$
cm$^{-3}$ (Bonamente, Lieu \& Mittaz \cite{blm2001}) a cooling
time of about 150 Myr.

Although strong evidence for cooling flows has been found in
low-resolution X-ray imaging and spectra, high-resolution
observations with the Reflection Grating Spectrometer on the
XMM-Newton satellite show inconsistencies with the simple cooling
flow models. The main problem is the lack of the emission lines
expected from gas cooling below $1-2$ keV \footnote{Several
solutions to this problem have been investigated, such as
heating, mixing, differential absorption and inhomogeneous
metallicity (Fabianet al. \cite{fmnp2001}), but each of these
hypotheses still remain under debate (Molendi \& Pizzolato
\cite{mp2001}). In particular, in the case of the Virgo cluster,
a multiphase gas model gives a better description of the available
data than a single temperature model only if it is not
characterized by a broad temperature distribution, but by a
narrow one (Molendi \& Pizzolato \cite{mp2001}). In the Virgo
core (up to a radius $\sim 10^{\prime}$), the gas temperature
ranges between 1 keV and 2 keV, so that the gas is not cold
enough to justify the observed EUV emission.}. Indeed,
observations with both the Hopkins Ultraviolet Telescope (Dixon,
Hurwitz \& Ferguson \cite{dhf96}) and FUSE (Dixonet al.
\cite{dixon2001a,dixon2001b}) found no significant far-UV line
emission from gas at $10^6$ K. \footnote{For the M87 galaxy Dixon,
Hurwitz \& Ferguson (\cite{dhf96}) found upper limits to far-UV
dereddened line emission of $4.84\times 10^{-6}$ erg cm$^{-2}$
s$^{-1}$ sr$^{-1}$ for the $O_{VI}$ line and $3.64\times 10^{-6}$
erg cm$^{-2}$ s$^{-1}$ sr$^{-1}$ for the $C_{IV}$ line. The
absence of a significant far-UV line emission, which is expected
from a warm gas component, is also confirmed by FUSE observations
not only for Virgo but also for the Coma cluster (Dixon et al.
\cite{dixon2001b}).} Also the observation of a large number of
clusters by the XMM-Newton satellite (Tamura et al.
\cite{tamura2001,0209332}) detected no warm gas component. All
these pieces of evidence seem overwhelming in the sense that a
thermal mechanism for the EUV excess can be ruled out.

Therefore, other mechanisms must be investigated as the source of
the EUV excess emission in clusters of galaxies. Indeed, Inverse
Compton (IC) scattering of Cosmic Background Radiation (CBR)
photons by relativistic electrons present in the ICM was proposed
to explain the EUV excess in the case of the Coma Cluster (Hwang
\cite{hwang}, Ensslin \& Biermann \cite{biermann98}, Ensslin,
Lieu \& Biermann \cite{elb99}).

Here, we will examine in detail the case of Virgo cluster,
referring in particular to the results of a re-analysis of
archival EUVE data for the central region of the cluster
performed by Bergh\"{o}fer, Bowyer \& Korpela (\cite{bbk2000}).

In Fig. 2 of the above cited paper, the azimuthally averaged
radial intensity profile of the background-subtracted EUV emission
in the central region of the Virgo cluster (centered in the M87
galaxy) is given, as well as the contribution of the low-energy
tail of the X-ray emitting hot ICM. Comparison between the two
curves shows the presence of a diffuse EUV excess which extends
up to a radius of $\sim 13 ^{\prime}$ (corresponding to $\sim 65$
kpc, for an assumed M87 distance $ D = 17$ Mpc), in the M87 halo
region \footnote{ The initial analysis of the same data claimed
to have detected EUV excess emission up to a distance of
$20^{\prime}$ (Lieu et al. \cite{virgo}).}.

The aim of the present paper is to show that IC scattering of CBR
and starlight photons by a population (hereafter denoted as
Population I) of mildly relativistic electrons with energy up to
a few hundred MeV is able to account for the EUV excess emission
observed from the central region of the Virgo cluster, centered
in M87 galaxy. As we will see in Section 2, Population I
electrons have to be described by a power law energy spectrum
with an energy cut-off at $E_{c} \sim 250$ MeV (or slightly
above) in order to avoid both a hard, non thermal, X-ray excess
(due to IC scattering of CBR photons) with respect to the upper
limit of $\simeq 4\times 10^{-12}$ erg cm$^{-2}$ s$^{-1}$ in the
2-10 keV energy band, set by observations (Reynolds et al.
\cite{rhfb}) and over-production of radio waves by synchrotron
emission (Owen, Eilek \& Kassim \cite{owen}).

The existence in the M87 galaxy of a highly relativistic electron
population (which we refer to as Population II) is well
demonstrated by the observations of large-scale radio emission in
the 10 MHz - 100 GHz band (Owen, Eilek \& Kassim \cite{owen}).
This radio emission is generally interpreted as synchrotron
radiation produced by relativistic electrons interacting with the
halo magnetic field. To b exact, taking a value $B \simeq 2.5~
\mu$G as typical for the M87 halo magnetic field, one gets an
electron energy range $\sim 1 - 170$ GeV. Population II electrons
have to be described with a power law energy spectrum with slope
$\alpha_e = 2.68$ in order to fit the observed radio power law
spectrum $\alpha_R = 0.84$. Electrons belonging to the same
Population II, but with energy in the range $0.7 - 1.7$ GeV, also
could give rise to a hard, non thermal X-ray emission in the 2 -
10 keV region (through IC scattering of CBR photons), but the
resulting power is always negligible with respect to the upper
limit quoted by Reynolds et al. (\cite{rhfb}).

Regarding the origin of Population I and II electrons, it is well
known that the main source of non-thermal energy in the M87 galaxy
is the active nucleus and jet. The current bolometric luminosity
of the inner active region is $\sim 10^{42}$ erg s$^{-1}$ -
mainly emitted in the radio band (Owen, Eilek \& Kassim
\cite{owen}) - although the kinetic power in the jet (which
matches with the nuclear power for accretion at the Bondi rate)
is $\sim 10^{44}$ erg s$^{-1}$ (Di Matteo et al. \cite{dimatteo}).
This is believed to be a lower limit for the total power
available for particle acceleration, since the central engine
would have had a cyclic activity with time scale of the order of
100-200 Myr and today be in a quiescent phase (Owen, Eilek \&
Kassim \cite{owen}; Corbin, O'Neil \& Rieke \cite{cor}).
Therefore, simple lifetime considerations indicate that the
available current energy content is at least $\sim 10^{60}$ erg.
On the other hand, in the past, the M87 galaxy could have
undergone a merging which would have made available a further
energy budget (as suggested for the Coma cluster by Blasi
\cite{blasi}). This energy budget could be stored by accelerated
protons that, being long-lived in the intergalactic medium (with
lifetimes of the order of a Hubble time) could be confined in the
galaxy and then released part of their energy to electrons by
Coulomb and/or hadronic interactions (see e.g. Ensslin
\cite{ensslin2002}) or by shock waves in plasma turbulence.

The observed radio luminosity requires that the power emitted by
Population II electrons is $\sim 10^{42}$ erg s$^{-1}$. Since the
cooling time of these electrons is dominated by synchrotron
losses with a time scale of $\sim 10^{15}$ s, one gets a
Population II energy budget of $\sim 10^{57}$ erg, much less than
the current available energy content. This implies that most of
the total energy made available by the M87 nucleus could be
stored in Population I electrons. If these electrons are
responsible for the EUV excess emission by IC scattering of CBR
and starlight photons, their power is about $10^{44}$ erg
s$^{-1}$, which multiplied by the IC cooling time $\sim 10^{17}$
s, gives a Population I energy content close to that available.

As far as the electron acceleration mechanism is concerned, it is
expected that Population II electrons have been directly
accelerated in the active nucleus and jet with high Lorentz
factors. Indeed, the two lobes in the observed radio map trace on
a large-scale the jet injection, keeping initially the direction
of collimated outflows from the core. Further diffusion by
magnetic field inhomogeneities may give rise to a diffuse
Population II component responsible for the overall radio
emission. It is also known that the active nucleus and jet in the
M87 inner region are responsible for inflating bubbles of
accelerated cosmic rays which rise through the cooling gas at
roughly half the sound speed (Churazov et al. \cite{cbkbf}).
Shock acceleration at the bubble boundaries could give rise to
Population I electrons.

Finally we note that the existence for Population I electrons of
an energy cutoff at $E_c \simeq $ 250 MeV may be justified by
considering that a time evolution for power-law spectrum injected
electrons is expected as a consequence of transport, diffusion
and cooling effects (see Petrosian \cite{petrosian} for a
detailed analysis for the Coma cluster).

In Section 2  we calculate the EUV source function for the IC
scattering by relativistic electrons on CBR and starlight photons
and the expected EUV excess radial profile towards the M87 galaxy.
In Section 3 we present our model results and address some
conclusions.

\section {Inverse Compton Scattering}

As stated in the Introduction, we assume that Population
I electrons account for
the large-scale EUV excess in M87 galaxy through
IC scattering of CBR and starlight photons.
As far as the electron spatial distribution is concerned,
despite evidence of granularity in the EUV excess map of M87
(Fig. 5 in  Bergh\"{o}fer, Bowyer \& Korpela \cite{bbk2000}),
here we adopt for simplicity a continuous radial profile
as representative of an
azimuthally averaged radial distribution of the observed structures.
Therefore, the electron density distribution $n_e(E_e,r)$ as a function of
the electron energy $E_{e}$ and the radial coordinate $r$,
in units of ${\rm cm^{-3}~ eV^{-1}}$, is
\begin{equation}
n_e(E_e,r) = \displaystyle{ \frac{K_I}
{\left[1+\left(\displaystyle{\frac{r}{r_e}}\right)^2\right]^{3\beta
/2}}
         ~g(E_e)},
\label{diste}
\end{equation}
where
\begin{equation}
g(E_e) = \left\{\begin{array}{ll}
\left(\displaystyle{\frac{E_e}{{\rm eV}}}\right)^{-\alpha_e}
~~~~~~~~~~~~~~~~~~~~{\rm for}~E_e <E_c \\ \\
\left(\displaystyle{\frac{E_e}{{\rm eV}}}\right)^{-\alpha_e}~e^{-(E_e-E_c)/E_s}
~~~{\rm for}~E_e > E_c~,
\end{array}\right.
\label{gE}
\end{equation}
so that the integral number density in the energy band $E_1-E_2$
can be calculated as
\begin{equation}
N_e(r) = \displaystyle{
         \int_{E_1}^{E_2} n_e(E_e,r)~dE_e }~.
\label{nde}
\end{equation}
Free parameters in eqs. (\ref{diste}) and (\ref{gE}) are the
constant $K_I$, the core radius $r_e$, the exponent $\beta$ of
the radial dependence, the electron spectral index $\alpha_e$,
the energy cutoff ($E_c$) and scale ($E_s$) parameters which, in
order to avoid over production of hard X-rays and radio waves,
turn out to be $E_c \simeq 250$ MeV and $E_s \simeq 125$ MeV,
respectively. The remaining parameters have to be determined in
order to fit the observed EUV excess radial profile in the energy
band $50 - 200$ eV. To this end we make use of the well-known
relation giving the IC scattered photon energy  (see e. g. Lang
\cite{lang})
\begin{equation}
E_{\gamma} = \frac{4}{3}~\left( \frac{E_e}{mc^2} \right)^2 ~<E_{ph}>
\label{eqno:eic}
\end{equation}
in terms of the average energy $<E_{ph}> \simeq 8/3~$KT$_{ph}$
of target photons when a black body spectral distribution
(at temperature $T_{ph}$) is assumed.

In our model the target photons are supplied by the CBR and
Starlight (S) photons, with temperature $T_{CBR} \simeq 2.7$ K and
$T_{S} \simeq 4400 $ K, \footnote{ The temperature $T_S$
associated with the starlight black body component can be
directly obtained from the photometry observations (see e.g.
Marcum et al. \cite{mof}) towards the M87 galaxy. From these
observations we derive that the absolute magnitudes in the B and
V bands are $M_B = -21.15\pm 0.07$ and $M_V=-22.08\pm 0.05$,
respectively. Consequently, the starlight black body temperature
turns out to be
\begin{equation}
T_S = \frac{7200~{\rm K}}{C_{B-V}+0.68}
\end{equation}
where $C_{B-V}=M_B-M_V =0.94\pm0.09$ is the M87 color index.
}
respectively.
According to eq. (\ref{eqno:eic}), Population I
electrons in the energy bands $125-250$ MeV and $3-10$ MeV
are involved in order to scatter the CBR and starlight photons,
respectively, into the EUV energy range $50-200$ eV.

The source function for EUV photon production through IC scattering
is given by (Lang \cite{lang}), in units of
${\rm cm}^{-3}~ {\rm s}^{-1}~ {\rm eV}^{-1}$, as
\begin{equation}
\begin{array}{ll}
Q_{IC}(E_{\gamma},r) = \displaystyle{
               \frac{c \sigma_T}{2} ~
               \left( \frac{mc^2}{{\rm eV}} \right) ^{1-\alpha_e}  ~
               \left( \frac{4}{3} \frac{ <E_{ph}>}{{\rm eV}} \right)
               ^{\frac{\alpha_e-1}{2}}  } \times \\ \\
~~~~~~~~~\displaystyle{
          N_{ph}(r) \frac{K_I}
{\left[1+\left(\frac{r}{r_e}\right)^2\right]^{3\beta /2}}
\times
\left(\frac{E_{\gamma}}{{\rm eV}}\right)^{-\frac{\alpha_e+1}{2}}
            }
~,
\end{array}
\label{eqno:icsource}
\end{equation}
where $N_{ph}(r)$
is the target photon number density profile, indicated as
$N_{CBR}$ and $N_S(r)$ for CBR and starlight photons, respectively.
Note that
for $E_{\gamma} < 200$ eV (corresponding to $E_e < E_c$)
the source function $Q_{IC}(E_{\gamma},r)$
shows a power law spectrum with
slope $\alpha_{EUV}=(\alpha_e+1)/2$.

As far as the average background photon density $N_{ph}(r)$
in eq. (\ref{eqno:icsource}) is concerned,
in the case of CBR photons it is clearly constant
($N_{CBR}\simeq 400$ cm$^{-3}$) within the M87 galaxy,
whereas for starlight photons it is obtained by the relation
\begin{equation}
N_{S}(r)  = \frac {J_{S}(r)} {<E_{S}> c}~,
\label{ns}
\end{equation}
in which the starlight energy flux $J_{S}(r)$, in units of
${\rm eV~s^{-1}~cm^{-2}}$,
is obtained from the integration of the local emissivity
$ {\cal L}_{S}(r^{\prime})$
on the M87 volume
\begin{equation}
J_{S}(r)  =
\int_{V} dV \frac  {{\cal L}_{S}(r^{\prime})}
{4 \pi |r-r^{\prime}|^2}~.
\end{equation}
We describe the local emissivity by the law
(Binney and Tremaine \cite{bt})
\begin{equation}
{\cal L}_{S}(r)  = \frac
{ {\cal L}_{S}(0) }
{ [1  + (r/r_{S}) ^ 2 ]^ {3/2} }~,
\label{distluce}
\end{equation}
which implies that the M87 starlight luminosity profile at impact
parameter $b$ is given by the modified Hubble profile
\begin{equation}
I_{S}(b)  = \frac
{ I_{S}(0) }
{ [ 1 + (b/r_{S})^2 ]} ~.
\label{distluce2}
\end{equation}
In the previous two equations the relation
${\cal L}_{S}(0) = I_{S}(0)/(2 r_{S})$ holds;
moreover $I_{S}(0)$ is connected to the total
luminosity through the relation
\begin{equation}
I_ {S}(0) = \frac
{ L_{S} }
{ \pi r_{S}^2 \log [1+(R_{M87}/r_{S})^2 ]}~.
\label{distluce3}
\end{equation}
Therefore, the whole photon density profile $N_{S}(r)$ due to
starlight is specified once the total luminosity $L_{S}$ and the
optical core radius $r_{S}$ are given. According to Tenjes,
Einasto \& Haud (\cite{teh}) and Zeilinger, M\o ller \& Stiavelli
(\cite{zms93}), we take $L_{S} = 1.07 \times 10^{11}~L_{\odot}$
and $r_S \simeq 566$ pc, respectively. For definiteness we have
taken the M87 galaxy size to be $R_{M87} \simeq 100$ kpc, although
our model results depend weakly  on the chosen value.

At this point we are ready to estimate the expected EUV excess radial
profile in the M87 galaxy.
By integrating the source function given
in eq. (\ref{eqno:icsource})
along the line of sight at impact parameter $b$
(neglecting internal absorption in M87),
we get the surface brightness profile (in units of
${\rm cm}^{-2}~{\rm s}^{-1}~{\rm eV}^{-1}$)
\begin{equation}
\Phi_{IC}(E_{\gamma},b) = 2 \displaystyle{\int_0^{\sqrt{R_{M87}^2-b^2}}}
              \frac{Q_{IC}(E_{\gamma},r)~r}{\sqrt{r^2-b^2}}~dr
~.
\label{eqno:phi}
\end{equation}
Correspondingly, the total power emitted in the energy band $E_1-E_2$
reads
\begin{equation}
L_{IC} = 2 \pi \int_0^{R_{M87}} db~b~
          \int_{E_1}^{E_2} dE_{\gamma}~\Phi_{IC}(E_{\gamma},b)~E_{\gamma}
~.
\end{equation}

The photon number count, in units of ${\rm arcmin^{-2}~s^{-1}}$,
 expected to be measured by the Deep Survey
telescope  aboard the EUVE satellite as a function of the
angular radius $\theta \simeq b/D$ results to be
\begin{equation}
{\cal N}_{EUV}(\theta)
= \displaystyle{\int_{E_1}^{E_2}} \Phi_{IC}(E_{\gamma},\theta D)
            e^{-\tau(E_{\gamma})}~A_{e}(E_{\gamma})dE_{\gamma}~,
\label{eqno:counts}
\end{equation}
where $[E_1-E_2] \simeq [50-200]$ eV is the sensibility energy
range of the DS instrument and $A_{e}(E_{\gamma})$ its effective
area given in Bowyer et al. (\cite{aeff}). In eq.
(\ref{eqno:counts}) we also take into account the galactic
absorption through the optical depth $\tau(E_{\gamma}) = N_H
\sigma(E_{\gamma})$, $N_H = 1.7 \times 10^{20}$ cm$^{-2}$ being
the Milky Way column density towards M87 (Hartman \& Burton
\cite{hb97}) and $\sigma(E_{\gamma})= [11~\sigma_{\gamma
H}(E_{\gamma}) + \sigma_{\gamma He}(E_{\gamma})] /12$ is
accounting for both the cross section of $\gamma H$ and $\gamma
He$ interaction, as parameterized by Morrison \& McCammon
(\cite{mmc83}) and Yan, Sadeghpour \& Dalgarno (\cite{ysd98}),
respectively.

\section{Results and conclusions}

\begin{figure}[htbp]
\vspace{7.7cm} \includegraphics{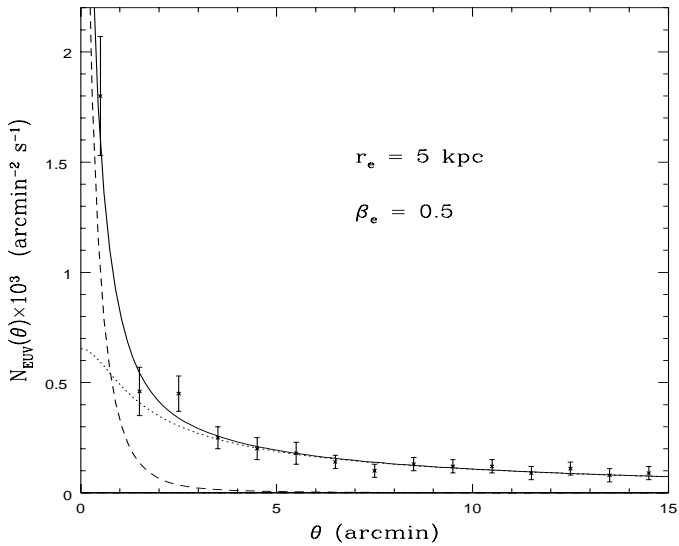}
 \caption{
The expected photon number count ${\cal N}_{EUV}$ is shown as a
function of the angular radius $\theta$, for the model giving the
best-fit (solid line) to the observational data. Dotted and
dashed lines represent the contribution from IC scattering of CBR
and starlight photons, respectively.
} \label{fit}
\end{figure}

Following the formalism discussed in Section 2, we fit the
available EUV excess data towards the M87 galaxy. For this purpose,
experimental points (given with the respective error bars in Fig. \ref{fit})
are derived from those given in
Fig. 2 in Bergh\"{o}fer, Bowyer \& Korpela (\cite{bbk2000}),
by subtracting the contribution
of the low-energy tail due to the X-ray emitting ICM.

Preliminarily, we have fitted the M87 observed EUV excess by
considering only the IC scattering of Population I electrons on
CBR photons  (Model I, with 11 d.o.f.). In this way, we obtain a
poor fit with $\chi^2=2.3$ by considering all the 15 observed
points (see the dotted line in Fig. \ref{fit}). \footnote{The
$\chi^2$ value gets reduced to $\chi ^2 \simeq 0.7$ if we do not
include the first observed point at $1^{\prime}$. Indeed, a
substantial fraction of the observed number count at $1^{\prime}$
may be the result of the core and jet activity (Bergh\"{o}fer,
Bowyer \& Korpela \cite{bbk2000}).}

Consideration of the IC scattering by the same electron
population also on starlight photons, distributed accordingly to
eqs. (\ref{ns}) - (\ref{distluce3}),  (Model II, with 9 d.o.f.),
allows us to substantially increase the expected photon number
counts from the innermost part of the galaxy, therefore obtaining
a much better confidence level $\chi ^2 \simeq 0.5$, considering
all the 15 points (see solid line in Fig. \ref{fit}). This
represents a significant improvement of the fit to all the
observed points since the outcome of the F-test implies that
Model II has to be preferred over Model I within a $3\%$
confidence level (the F-test value is 3.9).

Selected models with $\chi^2 < 1$ give a range of values for $r_e$
and $\beta$ given by $r_e = 5 \pm 1$ kpc and $ \beta = 0.5 \pm
0.1$. As far as the remaining two parameters $\alpha_e$ and $K_I$
- which are actually related by eq. (\ref{nde}) - our acceptable
$\chi^2 < 1$ fits give $ 2 < \alpha_e < 3$ and, correspondingly,
$10^4  \ut < K_I  \ut <  10^8$ eV$^{-1}$ cm$^{-3}$. The obtained
relative ratio between the EUV luminosity from IC scattering on
starlight and CBR photons is in the range of $2\%-6\%$. The
parameter $\alpha_e$ cannot be further constrained by EUV excess
data due to the absence of spectral measurements.

Recent experimental results (Reynolds et al. \cite{rhfb}) set an
upper limit of $4 \times 10^{-12}$ erg s$^{-1}$ cm$^{-2}$ for the
hard, non-thermal X-ray flux in the 2-10 keV energy band from the
M87 galaxy. Moreover, radio observations in the 10 MHz - 100 GHz
band give a total luminosity $L_R \simeq 5 \times 10^{41}$ erg
s$^{-1}$ (Herbig \& Readhead \cite{herbig}, Owen, Eilek \& Kassim
\cite{owen}). If we do not introduce an energy cutoff for
Population I electrons (see eq. (\ref{gE})), we would find both a
hard, non-thermal X-ray excess emission (in the 2-10 keV energy
band, due to IC scattering of CBR photons) and radio wave excess
by synchrotron mechanism, above the upper bound allowed by
observations.

In the following, as stated in the Introduction, we assume that
the radio emission is accounted for by Population II electrons
described by the same distribution law as in eq. (\ref{diste}) for
Population I electrons, but with different parameter values for
$K_{II}$, $r_e$, $\beta$, $\alpha_e$ and without energy cutoff.
\footnote{Indeed, Population I and II electrons are expected to
have different spatial distributions although the radio map of the
M87 halo shows some similar features with respect to the observed
distribution of the outer EUV excess (see Bergh\"{o}fer, Bowyer
\& Korpela (\cite{bbk2000}), Andernach, Baker, van Kap-herr \&
Wielebinski (\cite{abkw}) and Churazov et al. (\cite{cbkbf})).}
Here we recall that the characteristic frequency and the
differential luminosity of the ultrarelativistic synchrotron
radiation are given by (see e. g. Lang \cite{lang})
\begin{equation}
\nu = \frac{3}{4 \pi}\frac{eB}{mc}\left( \frac{E_e}{mc^2}\right)^2~
\label{eqno:syn}
\end{equation}
and
\begin{equation}
L_R(\nu) =  \frac{ \sqrt{3} e^3 C_R F_e }{8 \pi m c^2}
    \left( \frac{3 e}{4 \pi m^3 c^5}\right) ^{\frac{\alpha_e-1}{2}}
    B^{\frac{\alpha_e+1}{2}} ~ \nu^{-\frac{\alpha_e-1}{2}}~,
\label{eqno:nusyn}
\end{equation}
where $C_R \simeq (1.602 \times 10^{-12}~{\rm erg})^{\alpha_e}$ and
\begin{equation}
F_e =  \displaystyle{   \int_0^{R_{M87}}
\frac{K_{II}}
{\left[1+\left(\frac{r}{r_e}\right)^2\right]^{3\beta /2}}
~4 \pi r^2~dr}~.
\label{fe}
\end{equation}
If we take $\alpha_e \simeq 2.68$,
so that the predicted radio power law spectrum
$\alpha_R = (\alpha_e - 1)/2 \simeq 0.84$
is in agreement with the observed value,
and a value $B \simeq 2.5~\mu$G
for the magnetic field strength in the M87 halo (Dennison \cite{d80}),
there exists a set of acceptable values for
the free parameters in eq. (\ref{fe})
for which the obtained diffuse radio emission from the M87 galaxy
in the band 10 MHz -- 100 GHz
is $L_{R} \simeq 5 \times 10^{41}$ erg s$^{-1}$,
consistent with the observed value
(Herbig \& Readhead \cite{herbig},
Owen, Eilek \& Kassim \cite{owen}).

It is remarkable that the same set of values for Population II
electron parameters that fit radio observations give a hard,
non-thermal X-ray flux by IC scattering of CBR photons (by
electrons in the energy range 0.7 - 1. 7 GeV) in accordance with
the upper bounds given by Reynolds et al. (\cite{rhfb}) for the
M87 galaxy. \footnote{Of course, this estimate is obtained by
using the same formalism described in Section 2 for the EUV
excess by Population I electrons, but using in eq.
(\ref{eqno:icsource}) the appropriate Population II electron
parameter values.} More recently, the XMM-Newton observatory has
been pointed towards M87 galaxy with the aim of investigating if
non-thermal IC emission is present in the arms regions and if it
is linked to the power emitted by the radio jet. The upper limit
to the non-thermal X-ray emission has been found to be $\leq
4\times 10^{-14}$ erg cm$^{-2}$ s$^{-1}$ in the $0.5-8$ keV band,
representing less than $1\%$ of the flux from the thermal
components (Belsole et al. \cite{belsole}). Our model results
give a hard non-thermal X-ray flux from the inner $\simeq
5\arcmin$ region of M87 galaxy close to the above upper bound.
Clearly, our model results depend on the chosen parameters for
population II electrons which are not, at present, tightly
determined by radio observations. For example, by increasing the
Population II electron core radius (or decreasing the value of
$\beta$) we can dilute the X-ray excess (and also the radio
emission) within a wider region, leaving the same total power.

A few comments about the energy budget involved for the two
electron Populations are in order. Observations show that the
current bolometric luminosity of the M87 inner active region is
$\sim 10^{42}$ erg s$^{-1}$ - mainly emitted in the radio band
(Owen, Eilek \& Kassim \cite{owen}) - although the kinetic power
in the jet is $\sim 10^{44}$ erg s$^{-1}$ (Di Matteo et al.
\cite{dimatteo}). Simple lifetime considerations then indicate
that the available electron energy content is at least $\sim
10^{60}$ erg. However, the total power available for particle
acceleration may be larger since the central engine could have
had a cyclic activity with a time scale of the order of 100-200
Myr and today be in a quiescent phase (Owen, Eilek \& Kassim
\cite{owen}; Corbin, O'Neil \& Rieke \cite{cor}). Moreover, in
the past the M87 galaxy could have undergone a merging phenomenon
which would have made available a further energy budget,
initially stored by relativistic protons and later released
(partially) to electrons by Coulomb and/or hadronic interactions
or by shock waves in plasma turbulence (Ensslin
\cite{ensslin2002}).

Coming back again to our model results, we find that the energy
budget associated with the two electron Populations is about
$10^{60}$ erg (mostly in Population I electrons), which is always
less or approximately equal to the total energy associated with
thermal electrons. \footnote{The energy associated with thermal
electrons is estimated by assuming that the hot, virialized,
X-ray emitting gas has a solar metallicity composition and is
distributed following a $\beta$-model with $\beta \simeq 0.5$,
core radius $\simeq 1.6$ kpc and temperature $T_g \simeq 2.5$ keV
(Fabricant and Gorenstein \cite{fg}). Moreover, the gas central
number density is taken to be $\sim 3 \times 10^{-1}$ cm$^{-3}$
so that the total X-ray luminosity in the energy bands 0.2 - 4
keV and 2 - 10 keV turns out to be $\simeq 2.7 \times 10^{43}$
erg s$^{-1}$ and $\simeq 3 \times 10^{43}$ erg s$^{-1}$, in
accordance with the observed values (Forman and Jones \cite{fj}).}
This result represents an indication that the M87 galaxy is not
today in a quiescent (relaxed) phase.

However, further more accurate spectroscopic observations in both
the EUV and X-ray bands (especially towards the external region
of the M87 galaxy) are necessary in order to confirm the presence
of the two electron Populations invoked in the present paper.
Moreover, observations towards the other galaxy clusters for
which the EUV emission is still uncertain may establish whether
the existence of these two electron Populations is a common
feature in galaxy clusters or if it is particular to the central,
massive, cD galaxies like M87.

\acknowledgements{We thank the anonymous Referee for pointing out
some important issues.}

\end{document}